\documentstyle[prl,aps]{revtex}
\begin{document}
\title{Quantum Dynamics as a Stochastic Process}
\author{M. S. Torres Jr and J. M. A. Figueiredo\cite{email}}
\address{Universidade Federal de Minas Gerais - Dept. de F\'{i}sica\\
Caixa Postal 702 - Belo Horizonte - Brazil - 30.123-970}
\date{\today}
\maketitle
\pacs{02.50.Ey, 03.65.Ta, 03.65.Ca}

\begin{abstract}
We study the classical motion of a particle subject to a stochastic force.
We then present a perturbative schema for the associated Fokker-Planck
equation where, in the limit of a vanishingly small noise source, a
consistent dynamical model is obtained. The resulting theory is similar to
Quantum Mechanics, having the same field equations for probability measures,
the same operator structure and symmetric ordering of operators. The model
is valid for general electromagnetic interaction as well as many body
systems with mutual interactions of general nature.
\end{abstract}

\bigskip

\bigskip

Since their introduction in Physics stochastic processes became an
indispensable tool in the analysis of random phenomena. In many common
physical situations this mathematical theory arises as a natural description
language to such an extent that, rigorously speaking, we can't discard noise
effects from any realistic description of natural phenomena. On the other
hand Quantum Mechanics introduced probability concepts to physics without
explicit reference to any random phenomena, although it is believed that
this theory must generate a random vacuum field in order to explain more
subtle quantum processes. Thus it was quite natural that efforts in seeking
a stochastic nature to quantum phenomena have been presented, the pioneering
one being featured by Nelson \cite{nelson}. In the very beginning of the
Quantum Theory Planck \cite{Planck}, Einstein and Hopf \cite{Ein-Hopf} tried
to understand blackbody radiation in a statistical basis. Subsequently
Einstein and Stern \cite{Ein-Stern} developed a stochastic theory based also
on a random vacuum filed in order to explain low temperature corrections to
specific heat of the solids.

The advent of the Quantum Theory solved the intriguing problems issued by
the enormous sequence of experiments that opened the mysteries of the
microscopic world in the first two decades of the last century. The price
paid was the astonishing set of obscure epistemological foundations
presented by Quantum Theory. At that time such structural problems were a
scenario for deep discussions and controversies between Einstein and Bohr.
It is in EPR and Bell-like \cite{Hughes} experiments that researchers
usually look for evidence of a nontrivial indication that Quantum Mechanics
is in fact a fundamental theory. The successful predictions of Quantum
Mechanics have wakened physicists' epistemological demands in justifying its
basic axioms, and in many cases some authors were amazed by its intriguing
(epistemological) mysteries, which are believed to be inevitable and denying
any possibility of scientific reasoning on alternative paths.

Rescuing the original phenomenological view of the scientific thought of
Planck and Einstein, as presented in their first works, Boyer \cite{Boyer}
showed that relativistic invariance of the Einstein-Hopf vacuum leads to a
vacuum spectral density that is compatible with blackbody radiation law.
This may be a indication that a fluctuating vacuum field, if real, may
explain some observable phenomena only understood within the scope of the
Quantum Theory. Unfortunately Boyer%
%TCIMACRO{\UNICODE{0xb4}}%
%BeginExpansion
\'{}%
%EndExpansion
s work and its subsequent consequences proved to have no advantage over
Quantum Electrodynamics yet presenting some additional difficulties on many
calculations. Nevertheless, the idea of a real fluctuating vacuum everywhere
in the Universe should not be discarded since it provides a possible
empirical basis for a phenomenological theory compatible to Quantum
Mechanics without its epistemological inconsistencies and hopefully
providing additional, predictable phenomena.

In this work we present a theory that points toward this line. We study the
motion of a particle subject to a stochastic force of Wiener type \cite
{gardiner}, generating a specific kind of Brownian motion described by its
associated Fokker-Planck equation. The main ideas displayed here were
presented on an earlier work \cite{Figueiredo} where a deeper justification
for the stochastic process we consider is found as well as some consequences
not discussed this time. On the other hand we do make a generalization of
that work featuring a) the tridimensional case b) motion is subject to
electric and magnetic external fields and c) the many-particle case. As a
result we show that is possible to obtain a fully consistent physical theory
that is a proper generalization of Quantum Mechanics, based on phase space
dynamics, with full classical content. In addition the perturbative series
of the Fokker-Planck equation assumed here seems to be very efficient, since
it provides the correct quantum mechanical symmetric ordering of operators
coming from a full classical interpretation, without the need to calculate
redundant combinations of operators. In consequence a new term in the
expected values of forth order (and above) correlations shows up. A careful
calculation using Quantum Mechanics demonstrates the existence of this extra
term, never considered in conventional quantum mechanical calculations.

As stressed in the above considerations our picture is classical on its
epistemological nature. Nontrivial effects of noise is the main ingredient
on the kind of dynamical model we introduce now. Lets consider the motion of
a particle of mass $\mu $ subject to electric and magnetic fields. A
stochastic force is added to the associated Hamilton equations of motion
which are then given by 
\begin{eqnarray}
d{\bf x}\left( t\right) &=&\nabla _{p}{\cal H}dt\equiv \frac{1}{\mu }\left( 
{\bf p}\left( t\right) -\frac{e}{c}{\bf A}\left( {\bf x}\left( t\right)
,t\right) \right) dt  \label{Wienner} \\
d{\bf p}\left( t\right) &=&-\nabla _{x}{\cal H}dt+\sqrt{2\mu P}d{\bf W}%
\left( t\right)  \nonumber \\
{\cal H}\left( {\bf p},{\bf x},t\right) &=&\frac{1}{2\mu }\left[ {\bf p}-%
\frac{e}{c}{\bf A}\left( {\bf x},t\right) \right] ^{2}+V\left( {\bf x}%
,t\right)  \nonumber
\end{eqnarray}
where the subscript in the $\nabla $ operator stands for derivatives in the
corresponding variable, $P$ is a ''vacuum power'' and $d{\bf W}\left(
t\right) $ is a vectorial Wiener-type stochastic variable satisfying $d{\bf W%
}\left( t\right) ^{2}=dt$. As we will see the particle may gain energy from
the vacuum at a rate $P$. This apparently unphysical situation will be
carefully removed later when a proper perturbation theory will be developed,
leaving a consistent theory at the end of our reasoning. However we must be
aware of the fact that a real noise may exist. For example, since Universe
is expanding against a event horizon having a specific gravity, some energy
is necessary to change its curvature. Quantum mechanical calculations of
this effect leads to an effective ''vacuum temperature'' known as the
Gibbons-Hawking effect \cite{Gib-Hawk}. We may also invert the argument and
claim that vacuum effects supply the energy for the expanding Universe and
simultaneously supports quantum phenomena. Therefore it would not be a
surprise if, eventually, experiments come to confirm that, in fact, every
massive particle absorbs some energy from vacuum. This has in fact has been
considered by some cosmological models in order to explain the accelerated
expansion \cite{Turner}.

The motion in an electromagnetic field described by eqn$\left( \ref{Wienner}%
\right) $ demands the inclusion of a radiation reaction term that gives back
to vacuum part of the received power thus minimizing the effects of direct
observation of these fluctuations. We leave for a future work the inclusion
of this term since we believe it should not contribute to the main
conclusions of the present work. The pertinent point is that classical
orbits are very sensitive to stochastic perturbations {\em no} {\em matter}
the value of vacuum power. This means that even for infinitesimally small
value of $P$ the above equations admit a probabilistic description,
resulting in a Fokker-Planck equation for the associated probability
distribution. The procedure to get this equation from a Wiener source is
standard, essentially using general properties of Ito calculus \cite
{gardiner} along with statistical properties of the noise source. We
naturally arrive at a phase space description where classical trajectory
concept is lost and the associated phase space distribution function $\Phi
\left( {\bf x},{\bf p},t\right) $ satisfies 
\begin{equation}
\frac{\partial \Phi }{\partial t}=-\left( \nabla _{p}{\cal H}\right) .\left(
\nabla _{x}\Phi \right) +\left( \nabla _{x}{\cal H}\right) .\left( \nabla
_{p}\Phi \right) +\mu P\nabla _{p}^{2}\Phi  \label{FP1}
\end{equation}
Now ${\bf x}$ and ${\bf p}$ are no longer functions of time, meaning that we
can't assign a given trajectory to any point in phase space. All sample
trajectories are possible and, although real, they are not differentiable 
\cite{gardiner}, so a deterministic relationship between position and
momentum is not possible anymore. In short, the concept of a classical
deterministic motion is lost although particle reality, as well as its
trajectory, is not.

One general feature of this equation is that it is norm preserving. In fact
the time derivative of the norm is 
\[
\frac{d}{dt}\int \Phi \left( {\bf x},{\bf p},t\right) d\Omega =\int \left[
-\left( \nabla _{p}{\cal H}\right) .\left( \nabla _{x}\Phi \right) +\left(
\nabla _{x}{\cal H}\right) .\left( \nabla _{p}\Phi \right) +\mu P\nabla
_{p}^{2}\Phi \right] d\Omega 
\]
where $d\Omega =d^{3}{\bf x}d^{3}{\bf p}$ is phase space differential
volume. Direct integration by parts gives 
\[
\frac{d}{dt}\int \Phi \left( {\bf x},{\bf p},t\right) d\Omega =\int \left[
\left( \nabla _{x}.\nabla _{p}{\cal H}\right) \Phi -\left( \nabla
_{p}.\nabla _{x}{\cal H}\right) \Phi \right] d\Omega =0 
\]
This remarkable property is general for the kind of stochastic process we
consider here, being valid for any type of classical motion, regardless its
specific dynamics, since it is Hamiltonian-independent. This will be
considered later as a fundamental condition for defining the structure of
the probability measure supporting our theory. The following function, with
domain in phase space 
\[
H\left( {\bf x},{\bf p},t\right) \equiv \frac{1}{2m}\left( {\bf p}-\frac{e}{c%
}{\bf A}\left( {\bf x},t\right) \right) ^{2}+V\left( {\bf x},t\right) 
\]
cannot be confused with the (classical) particle's energy due to the
breakdown of trajectory reality but its average is an observable that
changes at a rate $P$ \cite{Figueiredo} 
\[
\frac{d}{dt}\left\langle H\right\rangle =\frac{d}{dt}\int \Phi \left( {\bf x}%
,{\bf p},t\right) H\left( {\bf x},{\bf p},t\right) d\Omega =P 
\]
and should be interpreted as the mean particle energy. Thus, as anticipated,
the particle gains energy at a rate $P$ that is proportional to noise
intensity. The (small) value of vacuum power would, if a real phenomenon, be
a thermodynamic property of the Universe, common to all existing elementary
particles. In this case, its measurable consequences can probably exist at
cosmological scales, but we will show that its stochastic effects are
dominant at microscopic ones, influencing the motion of small bodies and
giving rise to a dynamical process that, as we will see, resembles quantum
dynamics. This is the kind of motion we want to describe here. It shares
many common features with Quantum Mechanics but it is neither a Quantum
Theory nor a tentative task to give it some stochastic justification but a
full thermodynamic theory showing results that promises to be as precise as
Quantum Mechanics while still having the desired clearness of a
thermodynamic process.

Note that motion follows in mean classical laws and do not depend on vacuum
power. To see this observe that, although particle's velocity has only a
stochastic meaning, its average is a well defined time-dependent observable
given by $\mu \left\langle {\bf v}\right\rangle \equiv \left\langle {\bf p-%
\frac{e}{c}A}\right\rangle $ whose time derivative is given by 
\[
\mu \frac{d}{dt}\left\langle {\bf v}\right\rangle \equiv \frac{d}{dt}%
\left\langle {\bf p-\frac{e}{c}A}\right\rangle =\int \left[ \left( {\bf p-%
\frac{e}{c}A}\right) \frac{\partial }{\partial t}\Phi \left( {\bf x},{\bf p}%
,t\right) -\frac{e}{c}\frac{\partial {\bf A}}{\partial t}\Phi \left( {\bf x},%
{\bf p},t\right) \right] d\Omega 
\]
Direct manipulation of the Fokker-Planck equation $\left( \ref{FP1}\right) $
show that the particle follows in mean Newton's Law where a Lorentz force
term drives the motion. The result is 
\[
\mu \frac{d}{dt}\left\langle {\bf v}\right\rangle =\int \left[ -\nabla _{x}V-%
\frac{e}{c}\frac{\partial {\bf A}}{\partial t}+\left( {\bf p-\frac{e}{c}A}%
\right) \times \left( \nabla \times {\bf A}\right) \right] \Phi d\Omega
=\left\langle e{\bf E}+\left( {\bf p-\frac{e}{c}A}\right) \times {\bf B}%
\right\rangle 
\]
and is the stochastic version of Ehrenfest's theorem. This shows how vacuum
stochastic effects would be visible even for a vanishingly small value of
the vacuum power. Since motion suffers from random perturbations that smear
phase space trajectories no matter the value of noise intensity and since
the resulting average dynamics do not depend explicitly on its value a
nontrivial theory must survive even in the limit of zero power. This ''ghost
'' effect is a consequence of the known singular small-noise perturbation
expansion of the Fokker-Planck equation \cite{gardiner}. To solve eqn$\left( 
\ref{FP1}\right) $ we look for a perturbative schema that captures
explicitly these features.

The Fokker-Planck equation we have obtained is linear but does not have
constant coefficients implying that linear operations on it may not be
simple. This is particularly true for the spatial variable dependence due to
the arbitrariness of the external fields but the momentum variable has a
simpler linear dependence, allowing a tractable procedure in the Fourier
space. In this case we did arrive to a closed form of the transformed
equation where a very important feature turns out: due to its specific
dependency on the momentum variable, the resulting equation (in Fourier
space) is nonlinear on the scale conversion between adjoint spaces. Thus we
can't simply rescale and incorporate units in the transformed variables
because the physics changes. This means that only one value of the
conversion factor is compatible with experimental results. In the context of
the present work that's the way Planck's constant becomes important in
defining scales for microscopic phenomena. More explicitly, consider the
adjoint space of the momentum variable defined by 
\[
\varphi \left( {\bf x},{\bf y},t\right) =\int \Phi \left( {\bf x},{\bf p}%
,t\right) \exp \left( -i\frac{{\bf p.y}}{\hbar }\right) d^{3}{\bf p} 
\]
where $\hbar $ is Planck's constant. Here the $y$-space hosts modes of the
probability distribution associated to momentum which are waves having
wavelength of value $y\simeq \hbar /p$. Direct substitution of this
definition in eqn$\left( \ref{FP1}\right) $ results in the following
equation 
\begin{equation}
\frac{\partial \varphi }{\partial t}=\frac{\hbar }{i\mu }\frac{\partial
^{2}\varphi }{\partial y_{j}x_{j}}-\frac{1}{i\hbar }\frac{\partial U}{%
\partial x_{j}}y_{j}\varphi +\frac{e}{\mu c}\frac{\partial }{\partial x_{j}}%
\left( A_{j}\varphi \right) +\frac{e}{\mu c}y_{l}\frac{\partial A_{j}}{%
\partial x_{l}}\frac{\partial \varphi }{\partial y_{j}}-\frac{\mu P}{\hbar
^{2}}y_{j}y_{j}\varphi  \label{Fourier}
\end{equation}
\qquad where $U=V\left( {\bf x},t\right) +\frac{e^{2}}{2\mu c^{2}}A_{j}A_{j}$%
. As anticipated this equation is nonlinear in the constant $\hbar $ a fact
present even for zero vacuum power. In what follows, we expand in a Taylor
series in the variable ${\bf y}$ around the origin in order to get a
perturbation theory for this equation. This corresponds to a high kinetic
energy perturbation series. In \cite{Figueiredo} we did this (in one
dimension) by explicitly including a vacuum characteristic mode $k_{v}\equiv 
\sqrt{P/\left( \hbar c^{2}\right) }$ as a scaling factor in the expansion
which allows a non-singular phase space reconstruction. Since we intend to
take the limit $P\rightarrow 0$ later in the present work, a simpler but
rather singular approach (in phase space) will be developed here. We just
approximate the Gaussian kernel used in \cite{Figueiredo} by the unity,
albeit still preserving the diffusion term in eqn$\left( \ref{Fourier}%
\right) $. In one dimension, the limit of $k_{v}=0$ of the ref \cite
{Figueiredo} obviously coincides with ours. We believe that inclusion of
vacuum modes explicitly would be necessary if radiation reaction terms were
included in the Fokker-Planck equation in order to get a full thermodynamic
theory. Our task here is to show that a theory exists at sufficiently high
energy where direct vacuum effects have minor contribution to dynamics in
such a way the limit $P\rightarrow 0$ makes sense. In the present
representation in adjoint space we have developed an expansion close to the
origin in the variable ${\bf y}$ $\left( {\bf p}\rightarrow \infty \right) $
written as 
\begin{equation}
\varphi \left( {\bf x},{\bf y},t\right) =\sum_{l,m,n}\varphi _{l,m,n}\left( 
{\bf x},t\right) y_{1}^{l}y_{2}^{m}y_{3}^{n}  \label{expan}
\end{equation}
where the subscripts in ${\bf y}$ refer to its Cartesian components. When
inserted in eqn$\left( \ref{Fourier}\right) $ this results in a recursive
series for the coefficients $\varphi _{l,m,n}$, with only one of them
unknown since that equation involves only first order derivatives in ${\bf y}
$. At these high (kinetic) energies, vacuum modes have minor influence and
only appear explicitly after terms of order two. However due to the
existence of non-constant coefficients in eqn$\left( \ref{Fourier}\right) $,
even low order terms have a nonlinear dependence on Planck's constant,
leaving the expansion with nontrivial physics. Explicitly, we have 
\begin{eqnarray}
\frac{\partial \varphi _{l,m,n}}{\partial t} &=&\frac{\hbar }{i\mu }\left[
\left( l+1\right) \frac{\partial \varphi _{l+1,m,n}}{\partial x_{1}}+\left(
m+1\right) \frac{\partial \varphi _{l,m+1,n}}{\partial x_{2}}+\left(
n+1\right) \frac{\partial \varphi _{l,m,n+1}}{\partial x_{3}}\right] - 
\nonumber \\
&&-\frac{1}{i\hbar }\left[ \frac{\partial U}{\partial x_{1}}\varphi
_{l-1,m,n}+\frac{\partial U}{\partial x_{2}}\varphi _{l,m-1,n}+\frac{%
\partial U}{\partial x_{3}}\varphi _{l,m,n-1}\right] +  \nonumber \\
&&+\frac{e}{\mu c}\left( l\frac{\partial A_{1}}{\partial x_{1}}+m\frac{%
\partial A_{2}}{\partial x_{2}}+n\frac{\partial A_{3}}{\partial x_{3}}%
\right) \varphi _{l,m,n}+  \nonumber \\
&&+\frac{e}{\mu c}\left( l+1\right) \left( \frac{\partial A_{1}}{\partial
x_{2}}\varphi _{l+1,m-1,n}+\frac{\partial A_{1}}{\partial x_{3}}\varphi
_{l+1,m,n-1}\right) +  \nonumber \\
&&+\frac{e}{\mu c}\left( m+1\right) \left( \frac{\partial A_{2}}{\partial
x_{1}}\varphi _{l-1,m+1,n}+\frac{\partial A_{2}}{\partial x_{3}}\varphi
_{l,m+1,n-1}\right) +  \nonumber \\
&&+\frac{e}{\mu c}\left( n+1\right) \left( \frac{\partial A_{3}}{\partial
x_{1}}\varphi _{l-1,m,n+1}+\frac{\partial A_{3}}{\partial x_{2}}\varphi
_{l,m-1,n+1}\right) +  \nonumber \\
&&+\frac{e}{\mu c}\nabla .\left( {\bf A}\varphi _{l,m,n}\right) -\frac{\mu P%
}{\hbar ^{2}}\left[ \varphi _{l-2,m,n}+\varphi _{l,m-2,n}+\varphi _{l,m,n-2}%
\right]  \label{termogeral}
\end{eqnarray}
being implicit in this equation that the therms exist only for nonnegative
indices. We will show below that a consistent dynamical theory comes out for 
$P=0$, having properties common to Quantum Dynamics. For this, we collect
the following first two terms in the above equation which are independent of
the vacuum power $P$ \qquad 
\begin{mathletters}
\begin{eqnarray}
\frac{\partial \varphi _{0,0,0}}{\partial t} &=&\frac{\hbar }{i\mu }\nabla .%
{\bf J}+\frac{e}{\mu c}\nabla .\left( {\bf A}\varphi _{0,0,0}\right)
\label{tzero} \\
\frac{\partial J_{l}}{\partial t} &=&\frac{\hbar }{i\mu }\frac{\partial
T_{l,j}}{\partial x_{j}}-\frac{1}{i\hbar }\varphi _{0,0,0}\frac{\partial U}{%
\partial x_{l}}+\frac{e}{\mu c}\frac{\partial }{\partial x_{j}}\left(
A_{j}J_{l}\right) +\frac{e}{\mu c}\frac{\partial A_{j}}{\partial x_{l}}J_{j}
\label{tum}
\end{eqnarray}

where

\end{mathletters}
\begin{eqnarray*}
{\bf J} &\equiv &\left( \varphi _{1,0,0},\varphi _{0,1,0},\varphi
_{0,0,1}\right) \\
\text{{\sf T}} &\equiv &\left[ 
\begin{tabular}{lll}
2$\varphi _{2,0,0}$ & $\varphi _{1,1,0}$ & $\varphi _{1,0,1}$ \\ 
$\varphi _{1,1,0}$ & 2$\varphi _{0,2,0}$ & $\varphi _{0,1,1}$ \\ 
$\varphi _{1,0,1}$ & $\varphi _{0,1,1}$ & 2$\varphi _{0,0,2}$%
\end{tabular}
\right]
\end{eqnarray*}
are the probability current and the correlation tensor respectively. Another
set of equations involving $\varphi _{1,1,0}$ and its cyclic permutations as
well as the $\varphi _{1,1,1}$ term are also $P$-independent. Existence of
these $P$-independent coefficients is consequence of the high kinetic energy
character of the perturbative expansion. All coefficients are linked by a
recursive chain, so if we find a consistent way to calculate $\varphi
_{0,0,0}$ the entire series is solved. The first $P$-dependent term involves
time derivative of $\varphi _{2,0,0}$ (and its cyclic permutations) and
spatial derivatives involving order three terms. Thus it appears that $%
\varphi _{0,0,0}$ and ${\bf J}$ may be evaluated in a truly $P$-independent
way representing the ghost effect referred to above. This means that, even
for a very small vacuum source, the stochastic character of this dynamical
problem is not lost, thus preserving its probability foundation. Detectable
effects must exist and we have proved in \cite{Figueiredo} that (if real)
they take account of quantum phenomena of an elementary particle in its
nonrelativistic limit. Calculations made here give an improvement on the
mathematical structure of the theory, show the consistency in the
interaction with electromagnetic fields as well as the consistence in the
many particle case.

We initiate our analysis noting that $\varphi _{0,0,0}\left( {\bf x}%
,t\right) =\int \Phi \left( {\bf x},{\bf p},t\right) d^{3}{\bf p}$, meaning
the first coefficient is always real and its spatial integral equals the
time-independent norm in phase space. We get this result immediately by
integrating eqn$\left( \ref{tzero}\right) $ in the whole space. We introduce
the probability amplitude $\Psi \left( {\bf x},t\right) $ as $\varphi
_{0,0,0}\left( {\bf x},t\right) \equiv \left| \Psi \left( {\bf x},t\right)
\right| ^{2}$. The last argument demands that $\Psi \left( {\bf x},t\right) $
belongs to a particular ${\cal L}^{2}$ space ${\bf H}$ having the property 
\begin{equation}
\frac{d}{dt}\int \left| \Psi \left( {\bf x},t\right) \right| ^{2}d^{3}{\bf x=%
}\int \left( \Psi ^{\ast }\frac{\partial \Psi }{\partial t}+\Psi \frac{%
\partial \Psi ^{\ast }}{\partial t}\right) d^{3}{\bf x}=0  \label{dtnorm}
\end{equation}
which means $\left( \Psi ,i\partial _{t}\Psi \right) =\left( i\partial
_{t}\Psi ,\Psi \right) $ for all $\Psi \in {\bf H}$ where $\left( ,\right) $
stands for internal product in ${\cal L}^{2}$. It may be proved \cite
{Wheeler} that there is an Hermitean operator $\tilde{H}\left( {\bf x}%
,t\right) $ such that 
\begin{equation}
i\hbar \frac{\partial \Psi }{\partial t}=\tilde{H}\Psi  \label{hamilton}
\end{equation}
for every Hilbert space vector satisfying eqn$\left( \ref{dtnorm}\right) $.
Defining the (Hermitean) Hilbert space operators 
\begin{eqnarray*}
{\cal M} &\equiv &\frac{1}{2\mu }\left( \pi _{j}-\frac{e}{c}A_{j}\left( {\bf %
x},t\right) \right) \left( \pi _{j}-\frac{e}{c}A_{j}\left( {\bf x},t\right)
\right) +V\left( {\bf x},t\right) \\
{\bf \pi } &\equiv &\frac{\hbar }{i}\nabla
\end{eqnarray*}
it is not difficult to prove that for any $\Psi $ we have 
\[
\Psi ^{\ast }\left[ {\cal M},\pi _{j}\right] \Psi =-\frac{\hbar }{i}\frac{%
\partial U}{\partial x_{j}}\left| \Psi \right| ^{2}-\frac{e\hbar ^{2}}{2\mu c%
}\frac{\partial ^{2}A_{l}}{\partial x_{l}\partial x_{j}}\left| \Psi \right|
^{2}-\frac{e\hbar ^{2}}{\mu c}\Psi ^{\ast }\frac{\partial A_{l}}{\partial
x_{j}}\frac{\partial \Psi }{\partial x_{l}} 
\]
which upon integration in the whole space gives 
\begin{equation}
\int \Psi ^{\ast }\left[ {\cal M},\pi _{j}\right] \Psi d^{3}{\bf x=}-\frac{%
\hbar }{i}\int \frac{\partial U}{\partial x_{j}}\left| \Psi \right| ^{2}d^{3}%
{\bf x-}\frac{e\hbar ^{2}}{2\mu c}\int \frac{\partial A_{l}}{\partial x_{j}}%
\left[ \Psi ^{\ast }\frac{\partial \Psi }{\partial x_{l}}-\Psi \frac{%
\partial \Psi ^{\ast }}{\partial x_{l}}\right] d^{3}{\bf x}  \label{commut}
\end{equation}
We use this result in the space integration of eqn$\left( \ref{tum}\right) $
in order to get, after substitution for the potential energy term, 
\begin{equation}
\hbar ^{2}\frac{d}{dt}\int J_{l}d^{3}{\bf x=}\int \Psi ^{\ast }\left[ {\cal M%
},\pi _{l}\right] \Psi d^{3}{\bf x+}\frac{e\hbar ^{2}}{\mu c}\int \frac{%
\partial A_{l}}{\partial x_{j}}\left[ J_{l}-\frac{1}{2}\left( \Psi \frac{%
\partial \Psi ^{\ast }}{\partial x_{l}}-\Psi ^{\ast }\frac{\partial \Psi }{%
\partial x_{l}}\right) \right] d^{3}{\bf x}  \label{Jderiv}
\end{equation}
The freedom in defining $\Psi $ allows us to choose a specific functional
dependence for the probability amplitude. A natural choice would simplify eqn%
$\left( \ref{Jderiv}\right) $ in order to generate an equation for $\Psi $
and consequently closing the whole series. If we assume 
\begin{equation}
{\bf J}\equiv \frac{1}{2}\left( \Psi \nabla \Psi ^{\ast }-\Psi ^{\ast
}\nabla \Psi \right)  \label{current}
\end{equation}
we obtain a considerable simplification of eqn$\left( \ref{Jderiv}\right) $.
In addition the left side of this equation may be rewritten using eqn$\left( 
\ref{hamilton}\right) $ in order to get a form similar to the right side.
The result is 
\[
\hbar ^{2}\frac{d}{dt}\int {\bf J}d^{3}{\bf x=}\frac{\hbar ^{2}}{2}\frac{d}{%
dt}\int \left( \Psi \nabla \Psi ^{\ast }-\Psi ^{\ast }\nabla \Psi \right)
d^{3}{\bf x=}\int \Psi ^{\ast }\left[ \tilde{H},{\bf \pi }\right] \Psi d^{3}%
{\bf x} 
\]
in such a way that eqn$\left( \ref{Jderiv}\right) $ admits the (Hilbert
space) solution $\tilde{H}={\cal M}$. Using eqn$\left( \ref{hamilton}\right) 
$ a field equation for the probability amplitude is also obtained 
\[
i\hbar \frac{\partial \Psi }{\partial t}=\frac{1}{2\mu }\left( \frac{\hbar }{%
i}\nabla -\frac{e}{c}{\bf A}\right) ^{2}\Psi +V\left( {\bf x},t\right) \Psi 
\]
a condition that simultaneously satisfy eqn$\left( \ref{tzero}\right) $
completing the self consistency of the proposed solution. The current vector 
${\bf J}$ corresponds to the (classical) momentum of the particle averaged
over the momentum sector of the phase space 
\[
i\hbar {\bf J\left( x,t\right) =}\int {\bf p\Phi }\left( {\bf x},{\bf p}%
,t\right) d^{3}{\bf p} 
\]
so that all classical observables may be effectively calculated using a
similar rule \cite{Figueiredo}. Since the spatial average of $i\hbar {\bf J}$
is equal to the classical average of the momentum we have 
\[
\left\langle {\bf p}\right\rangle \equiv \int {\bf p\Phi }\left( {\bf x},%
{\bf p},t\right) d^{3}{\bf p}d{\bf ^{3}x=}\int \Psi ^{\ast }\left( \frac{%
\hbar }{i}\nabla \right) \Psi d^{3}{\bf x} 
\]
We have obtained a clean formulation of Quantum Mechanics with all axioms
and rules in a nice and phenomenological way as demanded by the traditional
scientific reasoning. In addition to the tridimensional formulation
presented here, the interaction with electromagnetic fields had the effect
of determining in a clever way the functional form of the probability
current because the vector potential explicitly couples to this current in
the perturbative expansion, a fact not possible in the one dimensional case.
Time derivative of eqn$\left( \ref{current}\right) $, along with
Schr\"{o}dinger equation, can be used in eqn$\left( \ref{tum}\right) $ in
order to get a explicit form of the correlation tensor. The result is 
\begin{equation}
T_{j,l}=\frac{1}{4}\left( \Psi ^{\ast }\frac{\partial ^{2}\Psi }{\partial
x_{j}\partial x_{l}}+\Psi \frac{\partial ^{2}\Psi ^{\ast }}{\partial
x_{j}\partial x_{l}}-\frac{\partial \Psi ^{\ast }}{\partial x_{j}}\frac{%
\partial \Psi }{\partial x_{l}}-\frac{\partial \Psi }{\partial x_{j}}\frac{%
\partial \Psi ^{\ast }}{\partial x_{l}}\right)  \label{tensor}
\end{equation}
and completes the determination of the two first terms in the series that do
not explicitly depend on the noise. Everything would work very well if
vacuum fluctuations are a real, detectable phenomena. In this case the
presumably small value of the vacuum power $P$ prevents us of its detection
using present experimental resolution, but since the two first terms in the
expansion used here for developing the theory are independent of its value,
a consistent dynamical model exists in the $P\rightarrow 0$ limit. On the
other hand, usual observables are at most second order on the dynamical
variables. Their averages calculated by the method presented here depends
only on the first two terms of the perturbative series and coincide with
those predicted by Quantum Mechanics \cite{Figueiredo} for any value of $P$.
Consequently their values are not affected by the vacuum power itself even
if it is real. In other words, no difference is noticeable between Quantum
Mechanics with all its axioms and mysteries and the thermodynamic theory
presented in this work in what concern a) field dynamics for $\Psi $ (the
Schr\"{o}dinger equation) and b) up to second order observables. The reason,
as anticipated, is the high kinetic energy limit used which prevents a
explicit dependence on vacuum effects although the stochastic nature of the
problem cannot be discarded even for zero vacuum power.

The very existence of other terms in the expansion show that the
Fokker-Planck equation represents a dynamical theory richer than Quantum
Mechanics at least in the nonrelativistic limit case. It is noticeable that
the probability distribution function $\Phi \left( {\bf x},{\bf p},t\right) $
cannot be confused with Wigner's function which, differently of our case,
does not satisfy the Fokker-Planck equation in the $P\rightarrow 0$ limit
where this equation assumes a Liouville-like form. Since our theory is
purely classical, expected values are calculated using classical variables
that are automatically converted in quantum-like operators by the
perturbative procedure we developed. In this aspect, Wigner's function
shares some commons properties with the distribution function given by the
Fokker-Planck equation. The particular case of the momentum angular
operators is interesting. We have 
\[
L_{l}=\varepsilon _{jkl}x_{j}p_{k}\Rightarrow \left\langle
L_{l}\right\rangle =\varepsilon _{jkl}\int x_{j}p_{k}\Phi d^{3}{\bf x}d^{3}%
{\bf p=}\int \Psi ^{\ast }\widetilde{L}_{l}\Psi d^{3}{\bf x} 
\]
where $\widetilde{{\bf L}}$ is the quantum mechanical angular momentum
operator. Thus each component of the averaged angular momentum coincides
with the quantum mechanical value. However the expected value of the squared
angular momentum differs from the usual quantum mechanical value. In fact we
have 
\begin{equation}
\left\langle {\bf L}^{2}\right\rangle =\int \left(
x_{j}x_{j}p_{k}p_{k}-x_{j}x_{k}p_{j}p_{k}\right) \Phi d^{3}{\bf x}d^{3}{\bf %
p=-}\hbar {\bf ^{2}}\int \left( x_{j}x_{j}T_{kk}-x_{j}x_{k}T_{jk}\right)
d^{3}{\bf x}  \label{L2 class}
\end{equation}
so by explicit calculation using eqn$\left( \ref{Jderiv}\right) $ we get 
\begin{equation}
\left\langle {\bf L}^{2}\right\rangle =\int \Psi ^{\ast }\text{{\bf $%
\widetilde{L}$}}^{2}\Psi d^{3}{\bf x}+\frac{3}{2}\hbar ^{2}  \label{L2}
\end{equation}
that is, the particle has an additional ''zero point'' squared angular
momentum equal to $\frac{1}{2}\hbar ^{2}$ per rotational degree of freedom.
The case of a free particle is sufficient to understand this result within
the framework of the present theory: while each component of the angular
momentum must be zero for that particle, stochastic effects prevent it to
describe a perfectly linear motion. Consequently, some motion in the plane
must exist, which demands the existence of a fluctuation in the total
angular momentum. By the same argument a rotor will also present a zero
point kinetic energy because $\left\langle E_{c}\right\rangle =\frac{1}{2I}%
\left\langle {\bf L}^{2}\right\rangle $, where $I$ is the moment of inertia
of the particle. Consider now a three dimensional harmonic oscillator; in
its fundamental state, the energy is $E_{0}=\frac{3}{2}\hbar \omega $ (both
in Quantum Mechanics and in the present stochastic model) and the total
angular momentum is zero, but our model predicts that a kinetic energy equal
to $\frac{1}{2I}\left\langle {\bf L}^{2}\right\rangle $ with $I=2\mu r^{2}$
and $r\equiv \sqrt{\hbar /\left( 2\mu \omega \right) }$ is present. It
appears that the stochastic model retains full consistence in the
interpretation of quantum phenomena.

The discrepancy to direct calculation using quantum theory displayed in the
eqn$\left( \ref{L2}\right) $ is removed when a fully symmetric quantum
operator is used. Consider the (classical) fourth order position-moment
correlation function $\chi 4_{ijkl}\equiv x_{i}x_{j}p_{l}p_{k}$. There are $%
24$ possible permutations of the associated quantum operators and many of
them are redundant permutations of commuting ones. Thus the complete full
symmetric quantum representation of $\chi 4$ is 
\begin{eqnarray*}
\widetilde{\chi 4}_{ijkl} &=&\frac{1}{6}\left( \widetilde{x}_{i}\widetilde{x}%
_{j}\widetilde{p}_{l}\widetilde{p}_{k}+\tilde{p}_{l}\tilde{p}_{k}\tilde{x}%
_{i}\tilde{x}_{j}\right) + \\
&&\frac{1}{12}\left( \widetilde{x}_{i}\widetilde{p}_{l}\widetilde{p}_{k}%
\widetilde{x}_{j}+\widetilde{x}_{j}\widetilde{p}_{l}\widetilde{p}_{k}%
\widetilde{x}_{i}+\tilde{p}_{l}\tilde{x}_{i}\tilde{x}_{j}\tilde{p}_{k}+%
\tilde{p}_{k}\tilde{x}_{i}\tilde{x}_{j}\tilde{p}_{l}\right) + \\
&&\frac{1}{24}\left( \widetilde{x}_{i}\widetilde{p}_{l}\widetilde{x}_{j}%
\widetilde{p}_{k}+\widetilde{x}_{j}\widetilde{p}_{l}\widetilde{x}_{i}%
\widetilde{p}_{k}+\widetilde{x}_{j}\widetilde{p}_{k}\widetilde{x}_{i}%
\widetilde{p}_{l}+\widetilde{x}_{i}\widetilde{p}_{k}\widetilde{x}_{j}%
\widetilde{p}_{l}\right) + \\
&&\frac{1}{24}\left( \widetilde{p}_{l}\widetilde{x}_{i}\widetilde{p}_{k}%
\widetilde{x}_{j}+\widetilde{p}_{l}\widetilde{x}_{j}\widetilde{p}_{k}%
\widetilde{x}_{i}+\widetilde{p}_{k}\widetilde{x}_{j}\widetilde{p}_{l}%
\widetilde{x}_{i}+\widetilde{p}_{k}\widetilde{x}_{i}\widetilde{p}_{l}%
\widetilde{x}_{j}\right)
\end{eqnarray*}
in such a way that after a simple but long calculation we get 
\begin{eqnarray}
\widetilde{\chi 4}_{ijkl}\Psi &=&\widetilde{x}_{i}\widetilde{x}_{j}%
\widetilde{p}_{l}\widetilde{p}_{k}\Psi +\frac{\hbar }{2i}\left( \delta
_{j,l}x_{i}+\delta _{i,l}x_{j}\right) \widetilde{p}_{k}\Psi +  \nonumber \\
&&\frac{\hbar }{2i}\left( \delta _{j,k}x_{i}+\delta _{i,k}x_{j}\right) 
\widetilde{p}_{l}\Psi -\frac{\hbar ^{2}}{4}\left( \delta _{i,l}\delta
_{j,k}+\delta _{i,k}\delta _{j,l}\right) \Psi  \label{coor4}
\end{eqnarray}
Since ${\bf L}^{2}=%
%TCIMACRO{\tsum}%
%BeginExpansion
\mathop{\textstyle\sum}%
%EndExpansion
\limits_{l,k}\left( \chi 4_{kkll}-\chi 4_{klkl}\right) $ the corresponding
symmetric operator is 
\[
{\bf \tilde{L}}_{N}^{2}\equiv 
%TCIMACRO{\tsum}%
%BeginExpansion
\mathop{\textstyle\sum}%
%EndExpansion
\limits_{l,k}\left( \widetilde{\chi 4}_{kkll}-\widetilde{\chi 4}%
_{klkl}\right) =%
%TCIMACRO{\tsum}%
%BeginExpansion
\mathop{\textstyle\sum}%
%EndExpansion
\limits_{l,k}\left( \widetilde{x}_{k}\widetilde{x}_{k}\widetilde{p}_{l}%
\widetilde{p}_{l}-\widetilde{x}_{k}\widetilde{x}_{l}\widetilde{p}_{k}%
\widetilde{p}_{l}\right) -\frac{2\hbar }{i}%
%TCIMACRO{\tsum}%
%BeginExpansion
\mathop{\textstyle\sum}%
%EndExpansion
\limits_{l}\widetilde{x}_{l}\widetilde{p}_{l}+\frac{3\hbar ^{2}}{2} 
\]
Expected value of ${\bf \tilde{L}}_{N}^{2}$ is given by $\int \Psi ^{\ast }%
{\bf \tilde{L}}_{N}^{2}\Psi d{\bf x}$, which is exactly the result displayed
in eqn$\left( \ref{L2 class}\right) $. This shows clearly that a
symmetric-ordered squared angular momentum operator also presents zero point
fluctuations as predicted by eqn$\left( \ref{L2}\right) $ which has a purely
classical interpretation. It appears that the perturbation expansion given
by eqn$\left( \ref{expan}\right) $ gives automatically the
symmetric-ordering correlation functions from the classical observables.
This means that quantum calculations using a classical construction without
axioms or rules is possible. This opens conditions for perturbative analysis
using higher order terms, which may come much easier with the above
formalism.

The many body problem can also be handled inside the present formalism. The
kind of noise considered here is uncorrelated, which means that the Wiener
process for this problem is just a set of Wiener variables for each
particle. The resulting Fokker-Planck equation has the same appearance shown
in eqn$\left( \ref{FP1}\right) $, the difference being the existence of an
additional index to label particles. We have found no differences on field
dynamics for $\Psi $ to Quantum Theory and, as above, all basic axioms and
rules come naturally. As expected correlation functions above forth order
present differences involving vacuum terms. At least in the fourth order a
symmetric-ordering is naturally obtained in a similar way worked out for the
one particle case.

The Fokker-Planck equation for the many particle case is 
\[
\frac{\partial \Phi \left( {\bf x},{\bf y},t\right) }{\partial t}%
=\sum_{\alpha =1}^{N}\left[ -\left( \nabla _{\alpha p}{\cal H}\right)
.\left( \nabla _{\alpha x}\Phi \right) +\left( \nabla _{\alpha x}{\cal H}%
\right) .\left( \nabla _{\alpha p}\Phi \right) +mP\nabla _{\alpha p}^{2}\Phi %
\right] 
\]
where greek symbols stands for particle labeling. Also ${\bf x}$ and ${\bf y}
$ stands for $\left\{ {\bf x}_{\alpha }\right\} $ and $\left\{ {\bf p}%
_{\alpha }\right\} $. Interactions include not only external fields but
mutual forces as well. The perturbative expansion analogous to eqn$\left( 
\ref{expan}\right) $ is written as 
\[
\varphi \left( {\bf x},{\bf y},t\right) =\rho \left( {\bf x},t\right)
+\sum_{\alpha }J_{\alpha i}\left( {\bf x},t\right) y_{\alpha }^{i}+\frac{1}{2%
}\sum_{\alpha \beta }T_{\alpha \beta ij}y_{\alpha }^{i}y_{\beta }^{j}+\frac{1%
}{6}\sum_{\alpha \beta \gamma }C_{\alpha \beta \gamma ijk}y_{\alpha
}^{i}y_{\beta }^{j}y_{\gamma }^{k}+... 
\]
and its coefficients satisfy equations similar to those found above. In
particular the correlation tensor $T_{\alpha \beta ij}$ has the following
form 
\[
T_{\alpha \beta ij}=\frac{1}{4}\left( \Psi ^{\ast }\frac{\partial ^{2}\Psi }{%
\partial x_{\alpha }^{i}\partial x_{\beta }^{j}}+\Psi \frac{\partial
^{2}\Psi ^{\ast }}{\partial x_{\alpha }^{i}\partial x_{\beta }^{j}}-\frac{%
\partial \Psi ^{\ast }}{\partial x_{\alpha }^{i}}\frac{\partial \Psi }{%
\partial x_{\beta }^{j}}-\frac{\partial \Psi }{\partial x_{\alpha }^{i}}%
\frac{\partial \Psi ^{\ast }}{\partial x_{\beta }^{j}}\right) 
\]
resulting in the angular momentum the expression 
\begin{equation}
\left\langle {\bf L}^{2}\right\rangle =\sum_{\alpha }\int \Psi ^{\ast }\text{%
{\bf $\widetilde{L_{\alpha }}$}}^{2}\Psi d^{3}{\bf x}+N\frac{\hbar ^{2}}{2}%
\frac{D\left( D-1\right) }{2}  \label{L2N}
\end{equation}
where $D$ is the spatial dimensionality of the problem. As before, the
additional vacuum term is common to particle-particle correlations above
fourth order and can be understood if symmetric-ordering of the associated
quantum operators is assumed. In this case, the generalization of the fourth
order operator displayed in eqn$\left( \ref{coor4}\right) $ is 
\begin{eqnarray*}
\widetilde{\chi 4}_{\alpha \beta \gamma \delta ijkl}\Psi &=&\widetilde{x}%
_{\alpha i}\widetilde{x}_{\beta j}\widetilde{p}_{\gamma l}\widetilde{p}%
_{\delta k}\Psi +\frac{\hbar }{2i}\left( \delta _{\beta \delta }\delta
_{j,l}x_{i}+\delta _{\alpha \delta }\delta _{i,l}x_{j}\right) \widetilde{p}%
_{\gamma k}\Psi + \\
&&\frac{\hbar }{2i}\left( \delta _{\alpha \gamma }\delta _{j,k}x_{i}+\delta
_{\beta \delta }\delta _{i,k}x_{j}\right) \widetilde{p}_{\delta l}\Psi -%
\frac{\hbar ^{2}}{4}\left( \delta _{\alpha \gamma }\delta _{\beta \delta
}\delta _{i,l}\delta _{j,k}+\delta _{\alpha \delta }\delta _{\beta \gamma
}\delta _{i,k}\delta _{j,l}\right) \Psi
\end{eqnarray*}
leading to an expression of the squared angular momentum equal to 
\begin{eqnarray*}
{\bf \tilde{L}}_{N}^{2} &\equiv &%
%TCIMACRO{\tsum}%
%BeginExpansion
\mathop{\textstyle\sum}%
%EndExpansion
\limits_{l,k}\left( \widetilde{\chi 4}_{kkll}-\widetilde{\chi 4}%
_{klkl}\right) =\sum_{\alpha ,\beta ,i,j}\left( \widetilde{x}_{\alpha i}%
\widetilde{x}_{\beta i}\widetilde{p}_{\alpha j}\widetilde{p}_{\beta j}-%
\widetilde{x}_{\alpha i}\widetilde{x}_{\beta j}\widetilde{p}_{\alpha j}%
\widetilde{p}_{\beta i}\right) - \\
&&-\frac{\hbar }{i}(D-1)\sum_{\alpha ,i}\widetilde{x}_{\alpha i}\widetilde{p}%
_{\alpha i}+N\frac{\hbar ^{2}}{2}\frac{D\left( D-1\right) }{2}
\end{eqnarray*}
whose expected value coincides with the classical result shown in eqn$\left( 
\ref{L2N}\right) $.

We see that a stochastic theory exists able to explain the results of
Quantum Mechanics. It is based on a phenomenological reasoning demanding
that vacuum fluctuations be real. We think that a phenomenological theory
has advantages over a purely axiomatic one because its foundations can
experimentally be tested and eventually enlarged. An axiomatic theory can be
extended owing to its results and by feeling only. Furthermore, a rational
interpretation of quantum phenomena is provided by the stochastic theory
enabling a comfortable epistemological basis. No new phenomena was predicted
within the limits of the nonrelativistic formulation presented here besides
additional terms in the correlation functions. However an explicit
calculation of Bell-like correlations may present new limits to quantum
measurements as well as shed some light on interpretation of entangled
states within the classical framework, possible in principle using the
stochastic approach.

Concluding we have shown that the limit $P\rightarrow 0$ of the
Fokker-Planck equation associated to a Wiener process describing the motion
of a charged particle in a electromagnetic field represents a well defined
stochastic process presenting all ingredients of Quantum Mechanics. This
means that, at least in the mathematical sense, there exists a stochastic
model that is exactly equivalent to Quantum Mechanics.

We have also shown that fourth order (classical) particle correlations are
compatible with symmetric ordering of quantum operators, opening
possibilities to get, in a simple way, higher order perturbation expansions
of quantum phenomena. Another advantage is the natural epistemological
scenario where all of the strange axioms of Quantum Mechanics come quite
naturally from a well defined phenomenological theory having a strong
thermodynamic appeal. This phenomenological foundation may be, at least in
principle, experimentally tested for vanishingly small vacuum power since,
in this case, we predict small differences to Quantum Theory. These
differences were shown in eqn$\left( \ref{termogeral}\right) $, where there
are additional terms involving the vacuum power $P$ that are expected to
give rise to correction terms in observables involving powers greater than
two of the particle's conjugate momentum. Anyway, we believe the present
model opens possibilities to a still better understanding of Nature inside
our classic rational approach to scientific reasoning.

\end{document}